\documentclass[fleqn,10pt]{wlscirep}
\usepackage{dcolumn}
\usepackage{amssymb,amsmath}
\usepackage{color}
\usepackage{epsfig}

\newcommand{\blue}{\textcolor{black}}
\title{\blue{Simple multi-wavelength imaging of birefringence: case study of silk}}

\author[1]{Reo Honda}
\author[1]{Meguya Ryu}
\author[2]{Jing-Liang Li}
\author[3]{Vygantas Mizeikis}
\author[4,5]{Saulius Juodkazis}
\author[1,*]{Junko Morikawa}
\affil[1]{Tokyo Institute of Technology, Meguro-ku, Tokyo 152-8550, Japan}
\affil[2]{Institute for Frontier Materials, Deakin University, Geelong, Victoria 3220, Australia}
\affil[3]{Research Institute of Electronics, Shizuoka
University, Naka-ku, 3-5-3-1 Johoku, Hamamatsu, Shizuoka 4328561, Japan}
\affil[4]{Swinburne University of Technology, John st., Hawthorn, 3122 Vic, Australia}
\affil[5]{Melbourne Center for Nanofabrication, Australian
National Fabrication Facility\\Clayton~3168, Melbourne, Australia}

\affil[*]{Correspondence: morikawa.j.aa@m.titech.ac.jp}

\begin{abstract}
\blue{A polarised light imaging} microscopy with an addition of liquid crystal (LC) phase retarder was implemented to determine \blue{the} birefringence of silk fibers with \blue{the} high $\sim 2~\mu$m \blue{spatial} resolution. The measurement was carried out with silk fiber (\blue{the optical} slow axis) and the slow axis of the LC retarder set parallel (a perpendicular alignment can also be used). The direct fit of the transmission data provides a high fidelity determination of birefringence, $\Delta n\approx 1.63\times
10^{-2}$ (with $\sim 2\%$ uncertainty) of the brown silk fiber (\emph{Antheraea pernyi}) averaged over the wavelength range $\lambda = (425 - 625)$~nm. By measuring retardance at four wavelengths it was pos\blue{s}ible to determine the true value of the birefringence of a thick sample when an optical path may include large number of wavelengths ($2\pi$ cycles in phase).  \blue{The n}umerical procedures and required hardware are described for the do-it-yourself assembly of the imaging polariscope at a fractional budget compared with commercial units.  
\end{abstract}
\begin{document}

\flushbottom
\maketitle
\thispagestyle{empty}
%%\email{sjuodkazis@swin.edu.au; morikawa.j.aa@m.titech.ac.jp}
%\tableofcontents
%``''
\section{Introduction}

%__________________________________Fig. 1
\begin{figure}[t]
\begin{center}
\includegraphics[width=8.50cm]{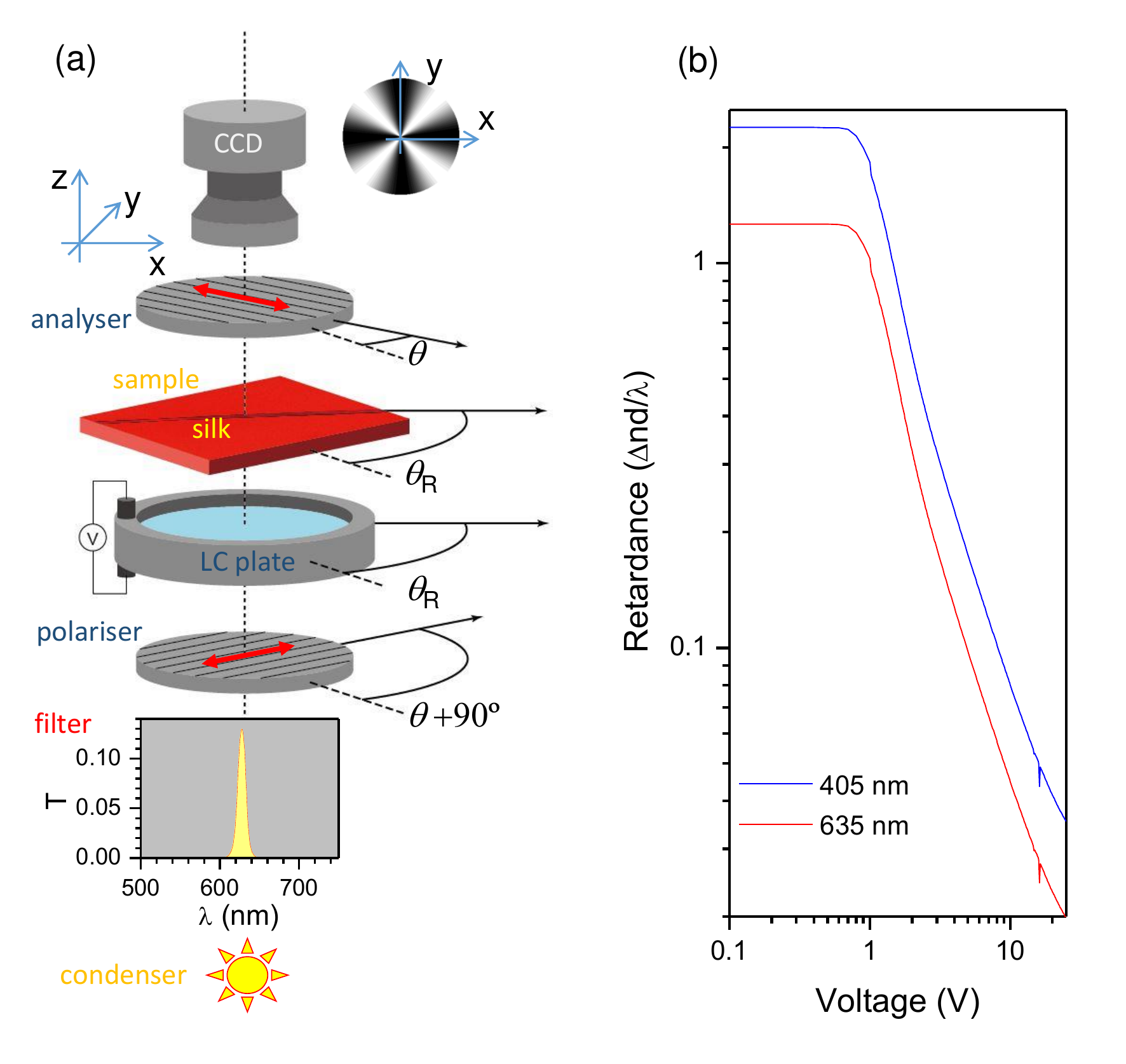}
\caption{(a) Schematic presentation of the assembled setup. \blue{The} Maltese cross \blue{(shown near CCD) is a polar plot of the mean intensity with variable $\theta-\theta_R$} (Eqn.~\ref{es}). A 10-nm-bandwidth filter (see, the actual spectral profile of transmission, $T$, plotted) was inserted to filter out the rest of the white light condenser illumination. Arrows mark \blue{transmission orientation of the} polariser and analyser. (b) Retardance vs voltage of liquid crystal cell (LCC1223T-A; Thorlabs) at two wavelengths \blue{calibrated by manufacturer.}} \label{f-lc}
\end{center}
\end{figure}
%__________________________________Fig. 2
\begin{figure}[tb]
\begin{center}
\includegraphics[width=9.0cm]{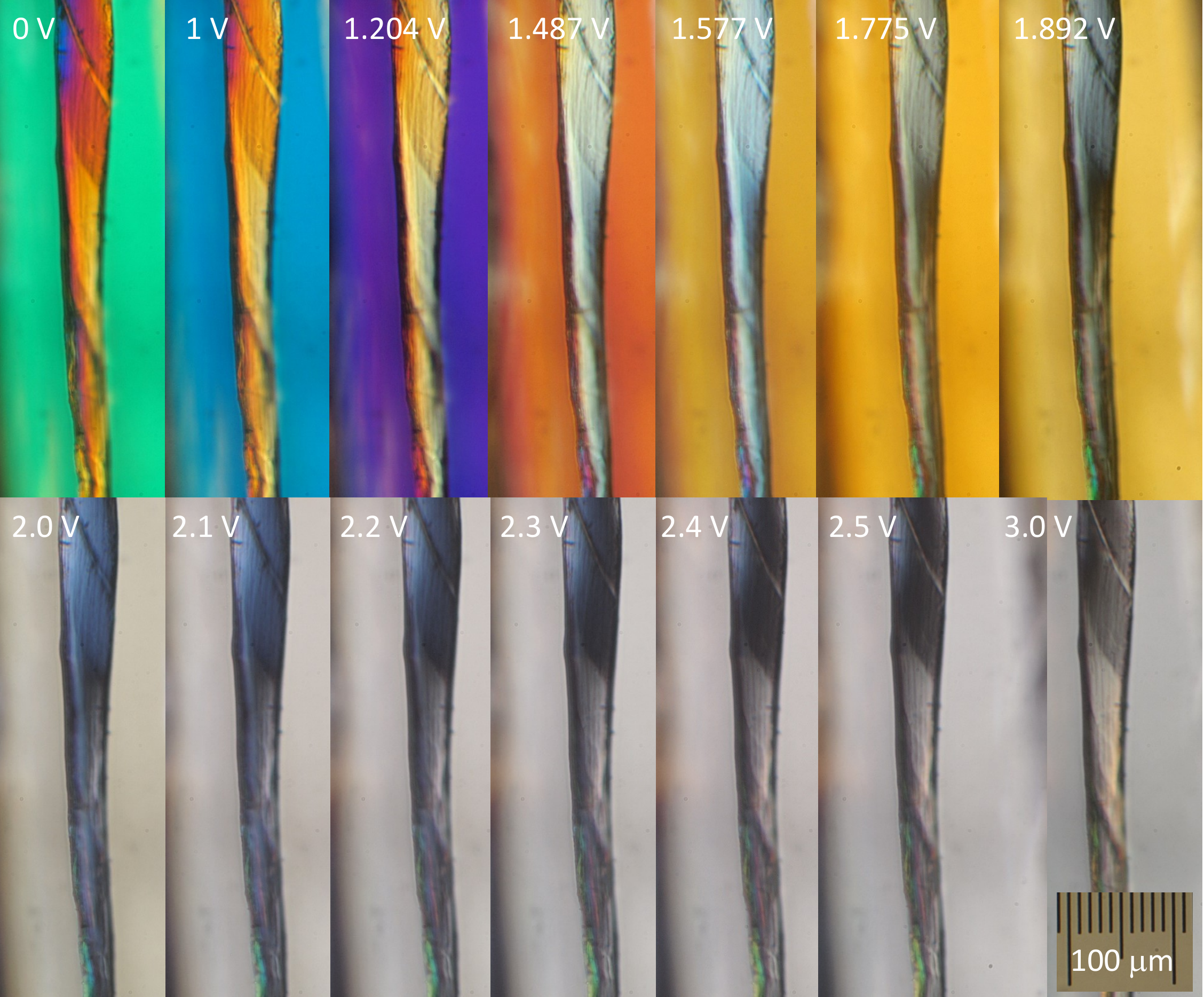}
\caption{Optical micro-images of white silk \blue{ (Bombyx mori)} through the crossed
polarizer-analyser under the white light illumination. Liquid crystal (LC) retarder voltage is marked. Slow-axis of the LC retarder was at
$\theta=45^\circ$ and silk fiber was perpendicular to the
slow-axis of the LC-retarder.} \label{f-white}
\end{center}
\end{figure}
%__________________________________Fig. 3
\begin{figure*}[tb]
\begin{center}
\includegraphics[width=14.50cm]{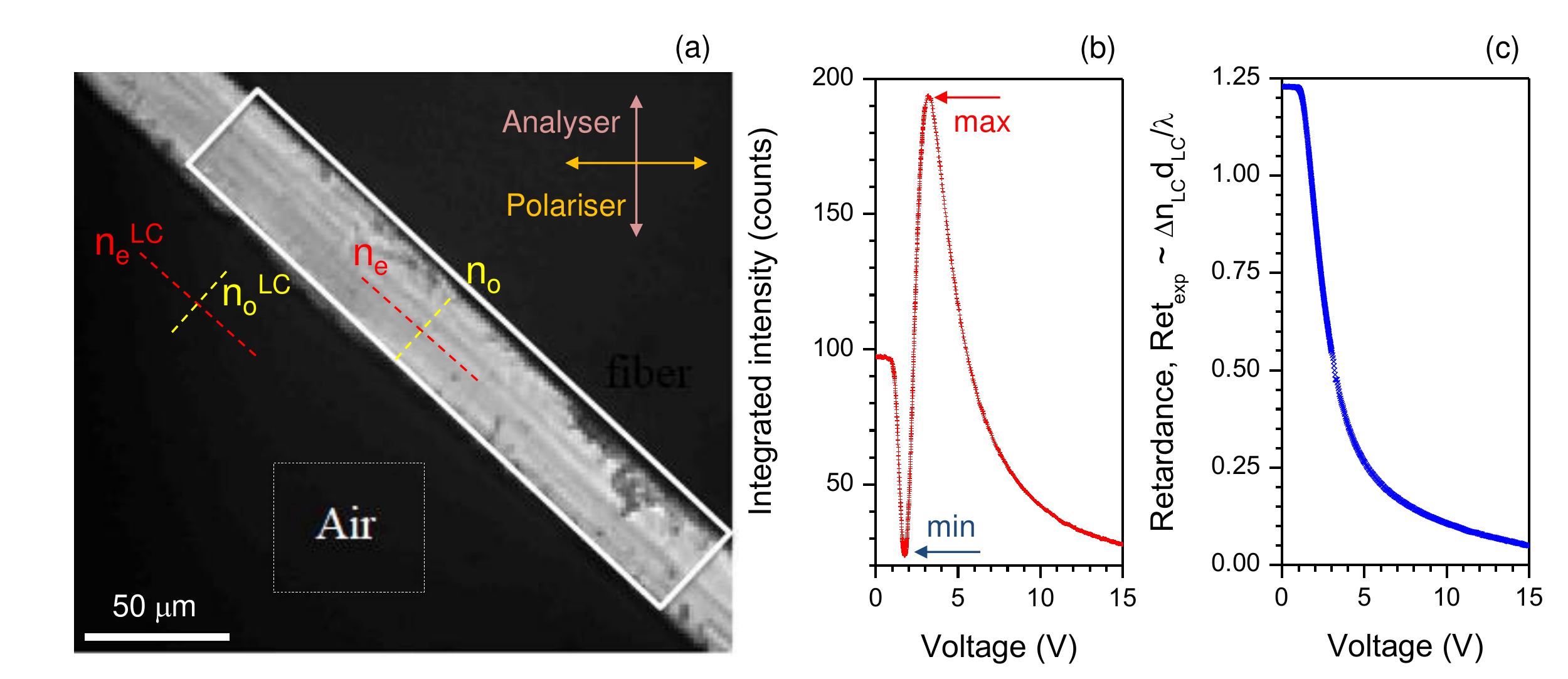}
\caption{(a) Optical image of a brown silk \blue{(Antheraea pernyi)} fiber at the liquid
crystal cell retarder voltage of 15~V. The band pass filter
at 635~nm wavelength was used for the white light condenser
illumination. The birefringence of silk is determined by the
difference of the extraordinary and ordinary refractive indices 
$\Delta n = n_e -n_o$. The dark region (out off the fiber sample) is where only the LC-retarder is between crossed polariser-analyser, marked as region of interest (ROI): ``Air''.  (b) The intensity integrated over selected ``Air'' (see, (a)) vs. the voltage (rms) 
of the retarder. (c) Digitised retardance of the LC cell $Ret_{exp}=\arcsin\sqrt{T_\theta}$, where $T_\theta = I_\theta/I_0$ (Eqn.~\ref{es}). Temperature of the LC-retarder was set
24.7-24.9$^\circ$C, exposure time 0.85~s. } \label{f-365}
\end{center}
\end{figure*}
 %__________________________________Fig. 4
\begin{figure*}[b!]
\begin{center}
\includegraphics[width=9.0cm]{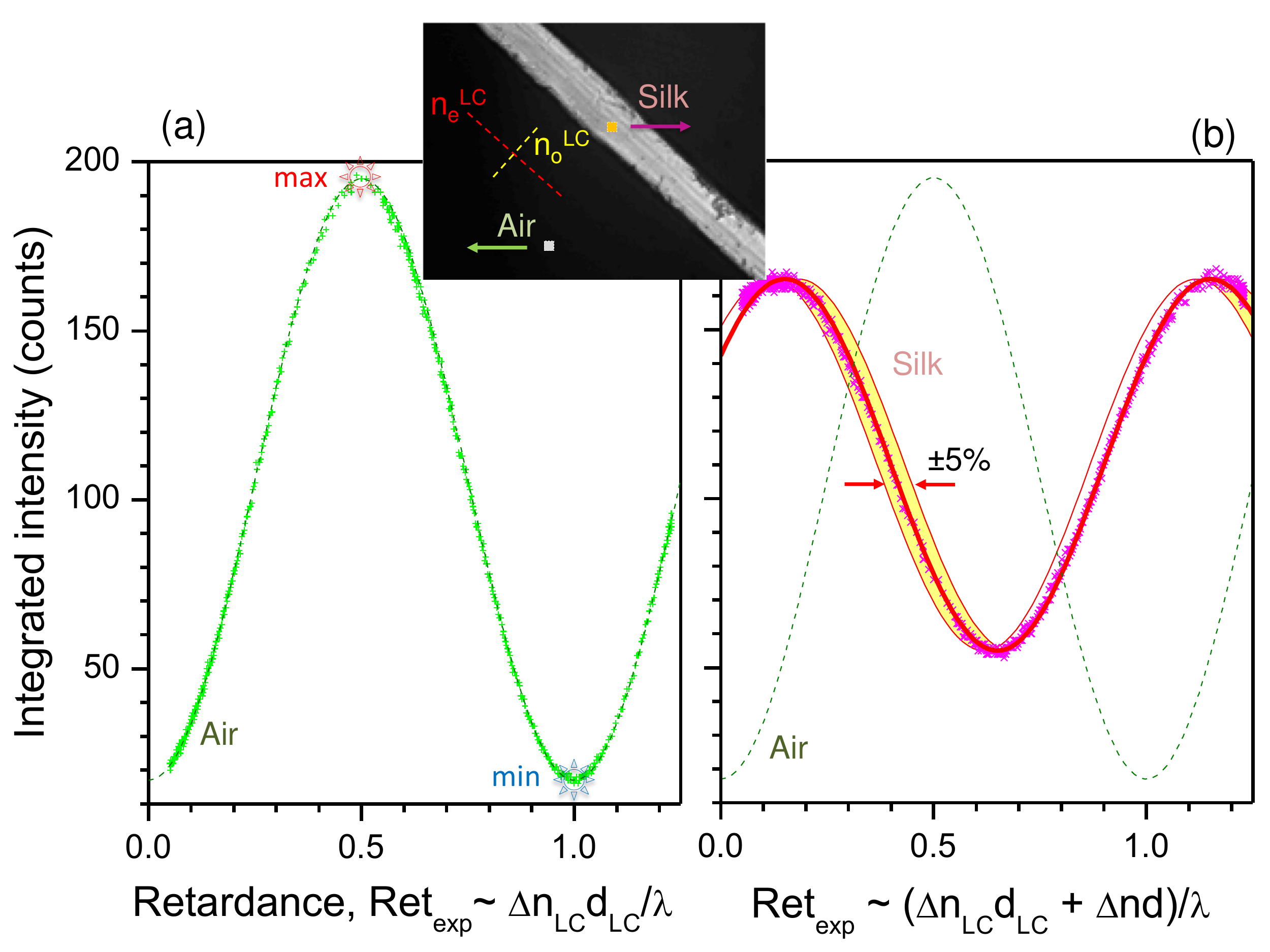}
\caption{Experimental measurement of the transmission. (a) Integrated
transmission intensity vs. retardance $Ret_{exp} = \arcsin\sqrt{T}$ for the marked region (see the inset) without silk fiber (marked square: ``Air''). Inset shows the image of the brown silk \blue{(Antheraea pernyi)} fiber at 15~V with highlighted regions
of the birefringence measurement $2\times 2$ pixels ($0.66\times 0.66~\mu$m$^2$); note, the resolution of the used objective lens was $\sim~2~\mu$m. (b) Integrated transmission through the silk fiber (dots) and the best fit (line). The fit function $fit(x)=a\sin^2(\pi[x + b])+o$, where the
best fit was achieved for selected \blue{amplitude, phase delay, and offset $a$, $b=\Delta nd$, and $o$, respectively,} and $\Delta n= 2.17\times 10^{-2}$ when $d = 30~\mu$m. } \label{f-retard}
\end{center}
\end{figure*}

Optical imaging of metasurfaces for definition of the engineered birefringence and its orientational pattern is gaining interest due to capability of direct evaluation of the fabrication quality and phase retardance in a fast growing field of flat optical elements~\cite{Capasso}. Determination of the slow-axis orientation and retardance can be made for an arbitrary sample using transmission polariscopy~\cite{abrio}. This principle was commercially implemented as a side-port addition onto a microscope (Abrio). However, a wider use of this technique was hampered by a comparatively high price; moreover, \blue{production of the unit} was discontinued. There is a need for the \emph{in situ} monitoring of birefringence in complex micro-fluidic flows~\cite{fluid}  and cell division microscopy where optical detection of a cell division could be monitored in real time~\cite{Russell} using a simple instrumentation. Measurements of birefringence are highly required in microscopy and material science fields, however, quite expensive dedicated microscopes or bulky add-on microscopy units have to be used and usually works at fixed wavelengths~\cite{abrio}. \blue{Birefringence can be inferred from Stokes poliarimetry which is realised by different principles of phase delay or polarisation rotation, e.g., based on photoelasticity~\cite{Kemp} or using liquid crystal (LC) retarders~\cite{Shribak1,Oldenbourg1}. Crossed polariser-analyser setup with a rotating quarter-waveplate compensator was used for determine a 3D orientation maps of birefringent fiber structures in a brain tissue~\cite{Zilles}. When waveplates are utilised together with rotating elements and lock-in amplifiers, setups of high sensitivity and resolution can be realised, however, they become bulky, complex and, frequently, wavelength specific.} 
% Miniaturisation, simplification, and spectrally broad functionality are strongly required. 

Simplification of \blue{the optical retardance} measurement at \blue{a} broader spectral range \blue{is} still strongly required especially for the flat optical elements and bio-materials with high orientation anisotropy and domain structure. For example, birefringence of silk is usually measured by a shear interferometry~\cite{Fouda,Omenetto,Dicko}, which does not provide a high resolution imaging capability. Emerging optical applications of transparent wood~\cite{Berglund} needs better understanding of optical properties of the micro-tubular wood structure. Stress induced birefringence in crystals/glasses/polymers~\cite{Martynas,10oe8300,11ass5439}, volume phase transitions~\cite{00n178}, and complex topological structures for volumetric stress control~\cite{Venkataramania}, or patterning of absorbance in transparent materials~\cite{16apa194} all produce complex optical anisotropy which needs high resolution, $\sim\lambda$, imaging.    

A set of crossed polariser and analyser is used to reveal \emph{qualitatively} the birefringence, $\Delta n$, of a sample placed between them using optical imaging at \blue{the} selected wavelength, $\lambda$. \blue{The} retardance $\Delta n d$ \blue{is defined by sample's thickness  $d$ and birefringence}. For the \blue{\emph{quantitative}} determination of \blue{the} birefringence and optical retardance, transmi\blue{ttance} is measured (Fig.~\ref{f-lc}). \blue{The t}ransmittance, $T$, through the birefringent medium of thickness, $d$, when reflectance and absorbance are negligible for the crossed polariser and analyser is given by (Fig.~\ref{f-lc}(a)):
\begin{equation}\label{es}
T_\theta = I_\theta/I_0 = \sin^22(\theta-\theta_R)\sin^2(\pi\Delta n d/\lambda),
\end{equation}
\noindent where $I_{\theta,0}$ are the transmitted and incident intensities, respectively, \blue{$\theta$ is the angle between the transmission axis of analyser and the horizontal x-direction of the view field} and it is positive for the anti-clockwise rotation (looking into the beam), $\theta_R$ is the slow (or fast) axis direction with the slow axis~\cite{Born_Wolf} usually aligned to the main molecular chain or along the polymer stretch or a silk fiber
direction \blue{as} in this study. Equation~\ref{es} represents the Maltese cross pattern shown in Fig.~\ref{f-lc}(a). 

By placing a sample on \blue{a LC} cell retarder \blue{(Fig.~\ref{f-lc}(a))} which has an electrically controlled birefringence \blue{(Fig.~\ref{f-lc}(b))}, it is possible to determine the birefringence of the sample by
compensating it with an opposite sign at the chosen orientation of the LC-retarder. Transmission vanishes at the regions with zero birefringence (Eqn.~\ref{es}). This is the principle used, e.g., in Berek compensator where, instead of a liquid crystal, \blue{a} tilting of \blue{the} birefringent quartz plate is used.

Here, we use a simple LC-cell as a birefringence compensator for
determination of silk fiber birefringence. \blue{The} sample is placed directly on the LC-cell window \blue{and aligned with slow-axis of the LC retarder. No any sample nor retarder rotation was required during measurements}. \blue{By u}sing a standard microscopy imaging at the freely chosen wavelength, the birefringence of a single
strand silk fiber was determined with high fidelity \blue{and} resolution \blue{(which can be comparable with the wavelength $\lambda$ at tight focusing}. This method is applicable to measure birefringence of any transparent materials over the visible 400-800~nm spectral range \blue{determined by transparency of LC cell}. Data acquisition and analysis were fully automated using Labview and Matlab codes. \blue{Due to virtue of multi-wavelength measurement capability, the proposed method allows to determine birefringence even when the retardance has  $2\pi$ phase changes. }

\section{Method and samples}
\subsection{Polarisation change due to birefringence}\label{method}

\blue{The used setup is based on a linearly polarised light illumination of the sample (Fig.~\ref{f-lc}(a)). Correspondingly, a simpler Jones matrix calculus (as compared to a more general Mueller calculus) is applicable to calculate the evolution of E-field of light as outlined next.} The x-polarised (horizontally) incident light is defined by the
E-field Jones vector (Fig.~\ref{f-lc}):
\begin{equation}\label{e1}
E_H = \left(%
\begin{array}{c}
  1 \\
  0 \\
\end{array}%
\right).
\end{equation}
The analyser is crossed and transmits only y-polarised (vertically)
light. The corresponding Jones matrix is given by:
\begin{equation}\label{e2}
 J_V = \left(%
\begin{array}{cc}
  0 & 0 \\
  0 & 1 \\
\end{array}%
\right).
\end{equation}
A generic Jones matrix of the retarder with the phase delay, $\phi =
k\Delta n d$, wavevector $k=2\pi/\lambda$, and with slow-axis at
angle, $\theta$, \blue{with} respect to the x-axis is given:
\begin{equation}\label{e3}
J_R(\phi,\theta)=\left(%
\begin{array}{cc}
  \cos\frac{\phi}{2}+i\sin\frac{\phi}{2}\cos(2\theta) & i\sin\frac{\phi}{2}\sin(2\theta) \\
  i\sin\frac{\phi}{2}\sin(2\theta) & \cos\frac{\phi}{2}-i\sin\frac{\phi}{2}\cos(2\theta) \\
\end{array}%
\right).
\end{equation}
In experiments, the LC-retarder was inserted with
silk fiber (the sample) oriented in parallel to the slow-axis of
the LC-retarder (Fig.~\ref{f-lc}). The Eqn.~\ref{e3} is used for the LC-retarder, $J_{LC}(\phi,\theta)$. Fig\blue{ure}~\ref{f-lc} shows setup
and calibration curves of the retardance vs. voltage. The silk fiber (sample) contributes to the phase retardance as
$J_s(\phi_s,\theta_s)$, where $\theta_s$ is calculated from
x-axis ($\theta_s = 0$). The overall transmission through the setup (Fig.~\ref{f-lc}(a)) is then:
\begin{equation}\label{e4}
\blue{E_{t} = (J_V J_s J_{LC}) E_i} = \left(%
\begin{array}{cc}
  0 & 0 \\
  A & B \\
\end{array}%
\right)E_H = \left(%
\begin{array}{c}
  0 \\
  A \\
\end{array}%
\right),
\end{equation}
\noindent where $A=a_1+ia_2$ and $B=b_1+ib_2$ are given in \blue{the}
Supplement; \blue{such} simplification of the cumulative matrix is due to the $J_V$. Intensity \blue{of the} transmission image
detected on CCD (Fig.~\ref{f-lc}(a)) is then $I=AA^*$, where $A^*$
is the complex conjugate. Further simplification of the trigonometric expressions $A,B$ takes place at $\theta_{LC} = \theta_s = \pi/4$ and allows a simple calculation of \blue{the} intensity at each $[x,y]$ pixel $I(x,y)$. \blue{By} matching $I(x,y)$ with the experimentally measured retardance $Ret_{exp}\equiv \arcsin\sqrt{T}$ (Eqn.~\ref{es}) allows to access the retardance $\Delta n\times d$. \blue{When the length of birefringent region $d$ is known (measured independently), the birefringence $\Delta n$ at each pixel can be calculated.} 

\subsection{Samples and measurements}
 
For imaging, a Nikon Optophot-POL microscope with an Olympus LMP PlanFLN  objective lens with 20$^\times$ magnification and numerical aperture $NA = 0.4$ was used. The CCD images with VGA resolution $480\times 640$ pixels where \blue{captured} (CCD camera BU030C Toshiba teli) for processing at $N = 718$ number of points (voltage \blue{values} of the LC-retarder cell). The LC-retarder (LCC1223T-A, Thorlabs) was used with \blue{the} TC200 controller and temperature stabilizer LCC25; the latter was used for the quantitative determination of the retardance. \blue{Factory calibration of retardance vs. applied voltage at selected wavelengths was provided by vendor (Fig.~\ref{f-lc}(a)), however, we applied a different calibration procedure suitable for any wavelength selected by the bandpass filter.} Image acquisition at different LC-retardance (number of points $N=718$) was computer controlled and \blue{a} Matlab code was used for the final image analysis using protocol described in Sec.~\ref{method}.  

Two types of silk white (Bombyx mori) and brown (Antheraea pernyi) were used in this study. Both were degummed, i.e., a sericin cladding desolved as described earlier~\cite{17m356} and single strands were used for imaging. \blue{Both silk types have similar composition and structure, hence, birefringence~\cite{17m356}. Brown silk strands have on average a slightly larger diameter. Image acquisition was carried out at room conditions and took several minutes for $N=718$ frames. }

\section{Results and discussion}

%__________________________________Fig. 5
\begin{figure}[bt]
\begin{center}
 \includegraphics[width=9.0cm]{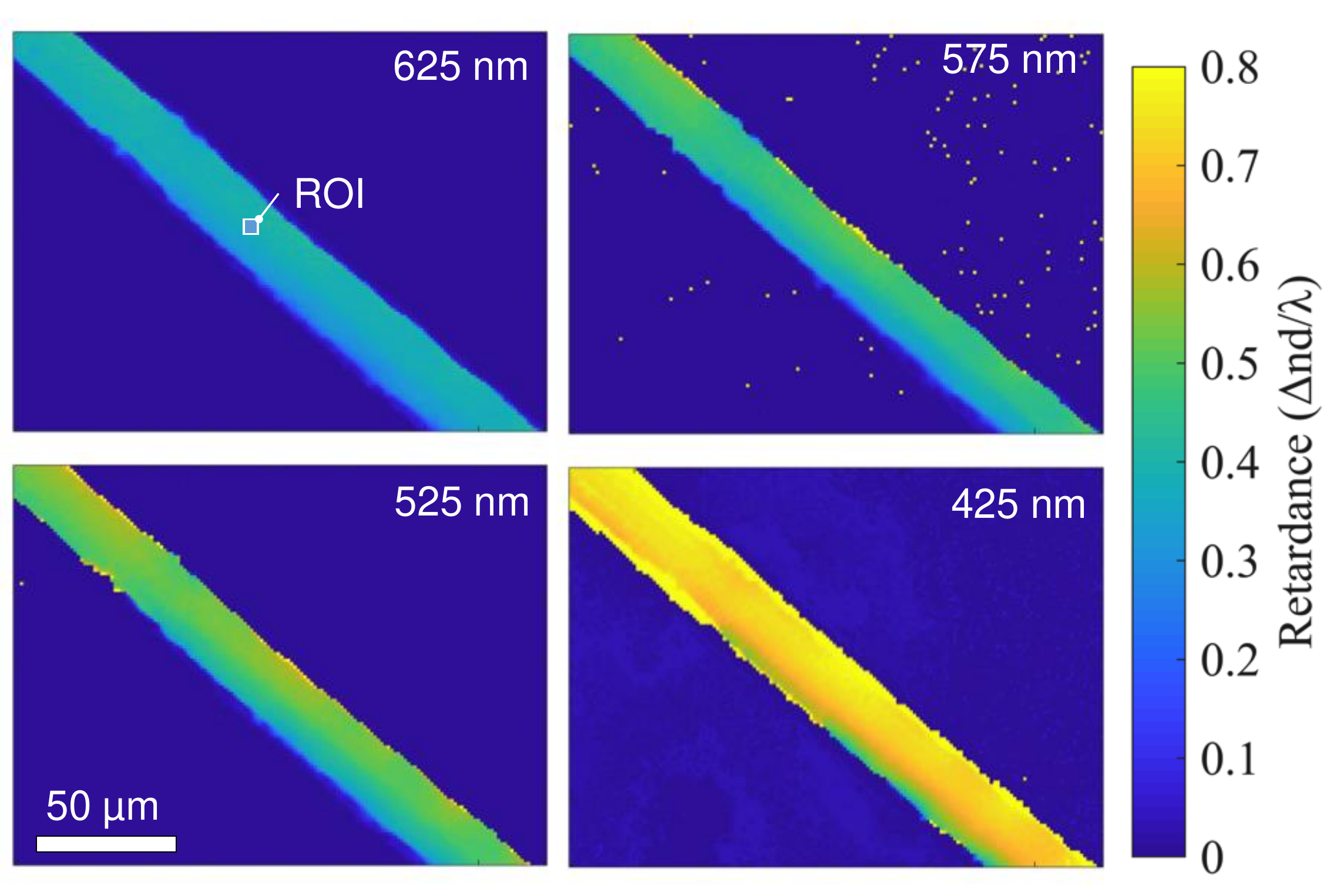}
\caption{Retardance $\Delta nd/\lambda$ map at four different wavelengths selected by interference filters (635, 575, 525, 425~nm) measured through the brown silk \blue{(Antheraea pernyi)} single strand. Since silk is naturally made from two strands,
after degumming an asymmetry of the strand is revealed (cross
section has a trapezoidal or triangular shape). The average $2\times 2$ pixels was used for numerical processing of the original VGA $480\times 640$ pixels CCD images. The optical resolution can be estimated as \blue{the} radius of Airy disk $w = 0.61\lambda/NA = 0.95~\mu$m for the used  $NA = 0.4$ objective lens. The silk fiber was placed on the LC-retarder which had the slow axis orientation parallel to the fiber. The ROI region was used to determine birefringence. }  \label{f-image}
\end{center}
\end{figure}
%__________________________________Fig. 6
\begin{figure}[bt]
\begin{center}
  \includegraphics[width=7.50cm]{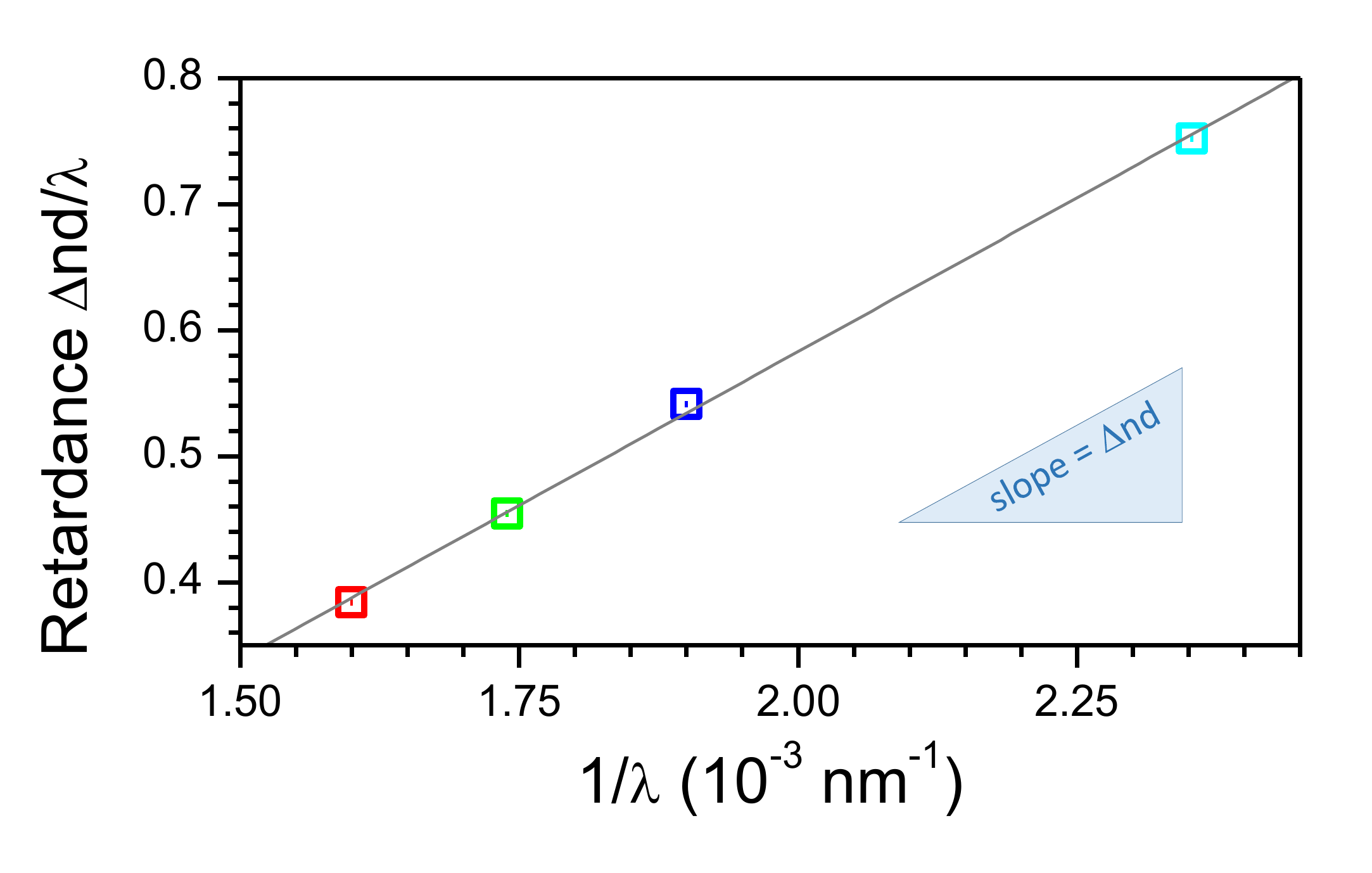}
\caption{Retardance $\Delta nd/\lambda$ vs $1/\lambda$  at four wavelengths (Fig.~\ref{f-image}) \blue{635, 575, 525, 425~nm presented by corresponding color markers}. Each point is average over ROI (see, Fig.~\ref{f-image}). Thickness of the brown silk \blue{(Antheraea pernyi)} fiber was $d \approx 30~\mu$m, which defines the birefringence $\Delta n\approx (1.63\pm 0.05)\times 10^{-2}$. \blue{The linear fit equation $y=p_1x+p_2$, coefficient and 95\% confidence interval are $p_1 = 487.5$ (from 450.6 to 524.5), $p_2=-0.3929$ (from -0.4639 to -0.3219), $R^2=0.9994$.}}  \label{f-line}
\end{center}
\end{figure}
%__________________________________Fig. 7
\begin{figure}[bt]
\begin{center}
  \includegraphics[width=8.50cm]{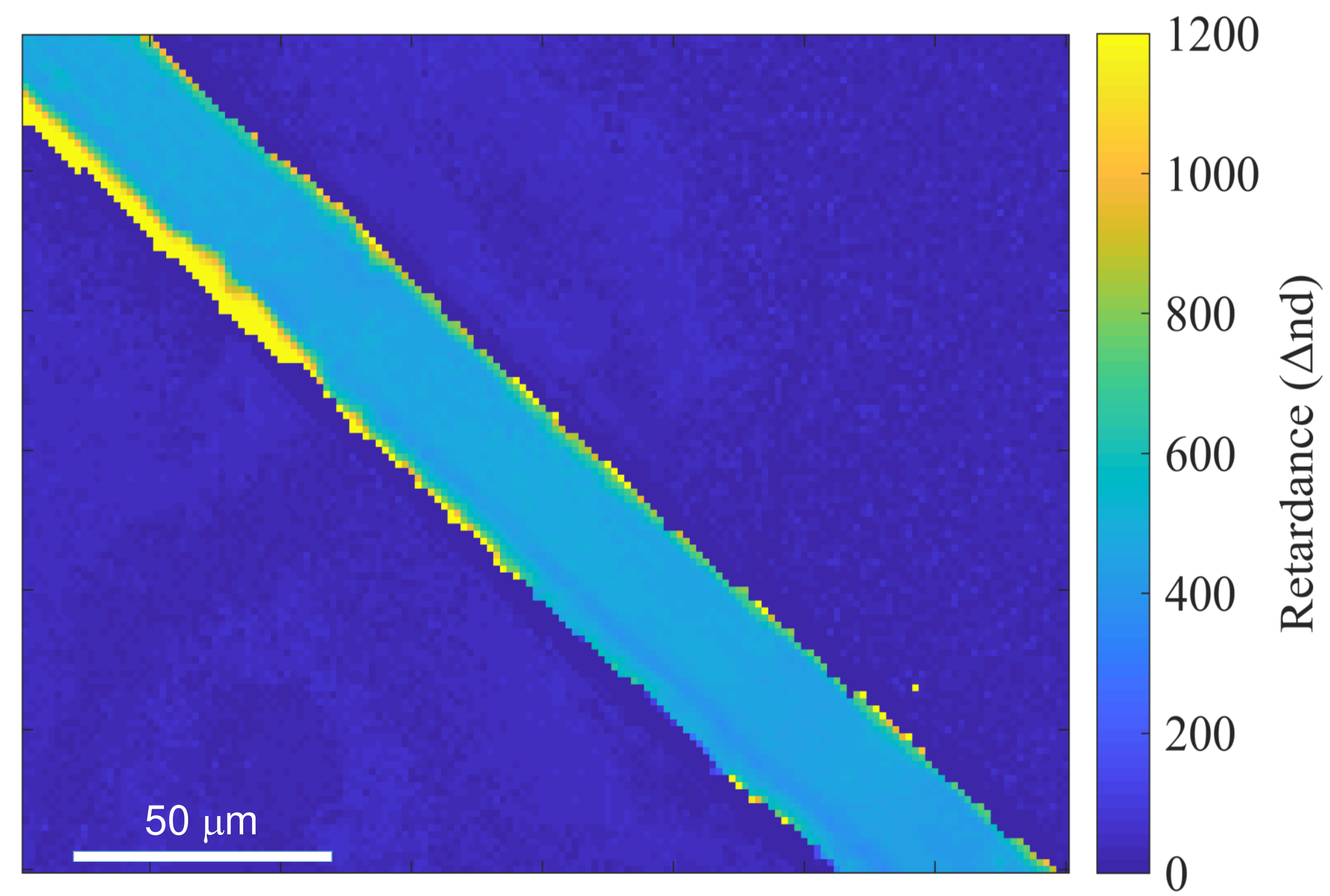}
\caption{Wavelength-averaged retardance ($\Delta nd$~[nm]) map (averaged over the four wavelengths; Fig.~\ref{f-image}). Thickness of the brown silk \blue{(Antheraea pernyi)} fiber was $d \approx 30~\mu$m, which defines average $\Delta n = (1.63\pm 0.05)\times 10^{-2}$ at the center of the silk fiber.}  \label{f-last}
\end{center}
\end{figure}

\blue{First, we show a \emph{qualitative} method of retardance imaging using silk fibers. Then, the \emph{quantitative} method is demonstrated using a simple LC retarder cell without waveplates (Sec.~\ref{method}).  }

\subsection{Qualitative imaging of retardance}

Phase retardance is used in polarisation microscopy to create
color contrast under a white light (condenser) illumination. It is useful for the qualitative distinction of regions with different birefringence (phase thickness) in the image. Figure~\ref{f-white} shows images of a single white silk \emph{Bombyx mori} fiber after deguming at different LC-retarder voltages. The slow-axis of the
LC-retarder was perpendicular to the silk fiber in order to
compensate the birefringence by decreasing retardance at larger
voltages (Fig.~\ref{f-lc}(a)). At the perfect compensation of birefringence, $\Delta n$, the dark
region is formed in the image as in the Maltese cross (Fig.~\ref{f-lc}(a)). The most dark region is changing its location in the image of the
fiber recognisable at large voltages (Fig.~\ref{f-white}). The fiber strands are, in fact, with a triangular or trapezoidal cross section which is causing a non uniform color appearance across the fiber.  However, only a qualitative estimate of the birefringence can be made using this method, even when imaging is carried at \blue{one} wavelength or at a spectrally narrow bandwidth.

\subsection{Quantitative imaging of retardance}

Using CCD imaging at different LC-retarder voltages (number of points $N = 718$) at different $\phi_{LC}$ values and by applying formulas Eqns.~\ref{e1}-\ref{e4}, it is
possible to determine the birefringence with high fidelity as described next. We \blue{used} a $\sim$10-nm-bandwidth filter \blue{to select a narrow spectral window from} the white light condenser illumination. \blue{S}ilk fiber \blue{was set} at 45~deg angle between the polarizer and analyzer (Fig.~\ref{f-365}). Also, the fiber was  parallel to the slow-axis of the LC-retarder $(\theta-\theta_R) = \pi/4$. A region of interest (ROI) ``Air'' was selected outside the silk fiber (Fig.~\ref{f-365}(a)) where only a reference retardance of the LC-cell was in the optical path. The $N = 718$ number of measurement points of transmi\blue{ttance was selected in equidistant steps of retardance} over the entire range of LC-retarder voltages \blue{as shown} in (b). To establish the relation between the average intensity on CCD (b), \blue{which is} proportional to the measured transmittance $T_{exp}=I_{Air}/I_0$, and to \blue{calculate} the retardance using Eqn.~\ref{es}, the intensity, $I_{Air}$, at the ``Air'' ROI (out of sample) was measured. The incident light intensity, $I_0$, was \blue{controlled by electrial current not} to cause saturation \blue{over} the entire $2\pi$ LC-retarder cycle. The minimum \blue{intensity} corresponded to the 0-phase (or $2\pi$) while the maximum to $\pi$ (Fig.~\ref{f-retard}(a)). Since $\sin^22(\theta-\theta_R) = 1$ by the selection of LC-retarder orientation $\theta_R =\pm\pi/4$, the reference retardance of the LC-cell is found from $T_{exp} = \sin^2(\pi\Delta n_{LC} d_{LC}/\lambda)$ (Eqn.~\ref{es}, where $n_{LC}, d_{LC}$ are the birefringence and thickness of the liquid crystal cell, respectively (Fig.~\ref{f-365}(c)).

The measured intensity averaged over ROI $2\times 2$~pixels (see the inset in Fig.~\ref{f-retard} from regions on the LC-retarder and on silk fiber are plotted in Fig.~\ref{f-retard}(a) and (b), respectively. Smaller ROIs were necessary due to a non uniform thickness of the fiber as described earlier and to test the smallest integration \blue{area; potentially the most noisy signal}. Since
the ``air'' region outside the silk fiber has no retardance
additional to \blue{that of} the LC cell, the transmittance follows
Eqn.~\ref{es}: $T_\theta = \sin^2(\pi\Delta n d/\lambda)$. Importantly, for the retardance corresponding to the half-wavelength $\Delta nd/\lambda = 0.5$ the transmittance \blue{h}as maximum (see arrow in
Fig.~\ref{f-365}(b)) and for the full wavelength $\Delta n
d/\lambda = 1$ it has a minimum (see arrow in Fig.~\ref{f-365}(b)). This was exactly what was expected and shows validity of the employed \blue{calibration} method. \blue{It was repeated for different set of bandpass filters defining different wavelengts. }

Figure~\ref{f-retard}(b) shows experimental\blue{ly measured} transm\blue{ittance} integrated over \blue{the $2\times 2$ pixels} ROI area on the silk fiber (rectangular box in the inset in (a)) vs. retardance of LC cell using the same procedure as used for \blue{the calibration of LC retarder shown in} (a). Even for a small number of the averaged pixels $2\times 2$, a high confidence fit by
$fit(x)=a\sin^2(\pi[x \pm b])+o$
was obtained with $a,o$ defining the amplitude and offset, $x=\Delta n_{LC}d_{LC}/\lambda$ is the retardance of LC-cell, and $b = \Delta nd/\lambda$ \blue{is determined by} silk with the sign conventions: ``+'' for the LC-cell orientation as shown in the inset of (a) and ``-'' for the one perpendicular to that. The phase of the $\sin$-wave was solely determined by the cumulative retardance through the LC-cell (``air'') and silk fiber $ \Delta nd$. The best fit for the LC-cell orientation shown in the inset of Fig.~\ref{f-retard}(a) was obtained with the ``+'' sign $fit(x)=110\sin^2(\pi(x+0.35))+55$ and corresponds to $\Delta n\approx 7.41\times 10^{-3}$ for the fiber thickness $d\approx 30~\mu$m \blue{(for simplicity silk fiber was assumed as a cylinder)}. The shaded region in Fig.~\ref{f-retard}(b) shows the $\pm 5\%$ change of $\Delta n$ \blue{around} the best fit \blue{value}. This shows \blue{qualitatively shows} that birefringence with $\pm 2\%$ \blue{difference can be distinguised; noteworthy, that a change of thickness equally affects the retardance as a change in birefringence.} \blue{It is noteworthy, that  neither the thickness nor orientation of the slow axis of fiber affect the offset of the sinusoidal curve in Fig.~~\ref{f-retard}(b). The variation of $d$ is the variation of retardance, therefore it changes the horizontal shift of the curve. Since the orientation of the slow axis is the first sinusoidal part of the Eqn.~\ref{es}, it affects only amplitude of the curve in Fig.~\ref{f-retard}(b). The only explanation of the offset is the depolarization of the light due to the scattering at the surface of the fiber sample.}

Retardance averaged over $2\times 2$~pixels was determined \blue{for} the entire image using the fitting shown in Fig.~\ref{f-retard}(b). It is presented in Fig.~\ref{f-image} for the four different wavelengths selected \blue{by} interference filters \blue{with} 10~nm bandwidth. \blue{The} 30-$\mu$m-thick silk fiber is effectively a half-waveplate at
525~nm wavelength. When thickness of the birefringent object is exceeding one wavelength ($2\pi$ in phase), there is an ambiguity in calculation of birefringence. To \blue{obtain the} exact $\Delta n$ value, retardance was measured at four wavelengths and fitted by  $(\Delta nd/\lambda \pm m)$ where $m = 0,1,2...$, as shown in Fig.~\ref{f-line}. A good linear fit was obtained for the retardance averaged over ROI (Fig.~\ref{f-image}) for $m=0$ plotted in Fig.~\ref{f-line}. For the central part of the fiber the birefringence $\Delta n\approx (1.63\pm 0.05)\times 10^{-2}$ was \blue{determined}. 

To obtain the map of the averaged retardance $\Delta nd$~[nm] over the spectral range from 425~nm to 625~nm, the same procedure as for the Fig.~\ref{f-line} was carried out for each $2\times 2$ pixels of the image at four wavelengths. From the slope of the linear fit, the $\Delta n$ was calculated (as in Fig.~\ref{f-line}) and is plotted in Fig.~\ref{f-last}. Edges of the silk fiber scattered light stronger which resulted in a \blue{higher detected} light intensity $T_\theta$ (Eqn.~\ref{es} ) and correspondingly up to twice larger effective retardance. Noteworthy, the thickness of silk fiber is smaller at the edges.   

% Retardance can be determined within a $\sim 2\%$ precision which corresponds to the $\sim 4.3\times 10^{-4}$ in birefringence with resolution down to 1-2~$\mu$m used in this study \blue{(defined by the $NA$ of the imaging optics and can reach $\sim\lambda$)}.

\section{Conclusions and outlook}

In summary, a simple LC-retarder addition to \blue{polarisation} microscopy provides a highly sensitive method to image birefringence as demonstrated for silk fibers. The proposed method relies on a large data set (sampling) of images obtained at different LC-retarder voltages (phase delays) used for  the best fit (a minimum of four images with $\pi/4$ LC-phase delays are required for the $\sin$-fit). It is shown that birefringence of silk $\Delta n \approx 1.6\times 10^{-2}$ was determined with an uncertainty of $\sim\pm 2$\% from \blue{the} area of just $2\times 2$~pixels \blue{at the center of $30~\mu$m-thick brown silk fiber} with \blue{a} high fidelity. Smaller number of images or integration over lager ROI areas can be flexibly applied to achieve a better spatial resolution or an average birefringence, respectively. Measurements at several wavelengths were \blue{made} to establish the absolute phase retardance. This is one of the strong features of the proposed method \blue{since most of} commercial microscopy-based realisations of birefringence measurements are usually carried out at one wavelength. \blue{T}he multi-wavelength measurement allows to extend retardance range beyond the $2\pi$ in phase.

The proposed method is much simpler and requires a fractional budget of $\sim\$2$k as compared with the established birefringence measurement microscopes. When the slow axis of the sample is unknown or it is changing orientation over the image area, the ax\blue{ial} alignment can be \blue{made} by an additional measurement at four points (the minimum number required for the fit) of the angular orientation of the sample and to implement calculations for the corresponding $A, B$ values Eqns.~\ref{es}~\ref{e4}. This functionality can be easily added. Also, an absorption anisotropy \blue{(diattenuation)} can be measured using transmission with adequately high resolution $\sim\lambda$ using this simple technique \blue{for} analysis of molecular alignment~\cite{Cruz}. 

% Since the dependence eqn.~\ref{es} is fitted by a $\sin$-curve, only four points of the transmission measurements suffice for the reliable fit (rather $N=718$ used here). This will further simplify and accelerate measurements.     
The proposed technique could also find application in bio-medical field
for cell monitoring and optical detection of cell division exploiting a new dimension - the birefringence - in addition of the usual set of the big-data dimensions of the lateral  $xy$-position of the cell, time, intensity and shape of the object. The enhanced light scattering at the edges would be beneficial for determination of the outline of the dividing cells. 

\subsection*{Acknowledgements}
\small{JM acknowledges partial support by a JSPS Kahenki Grant
No.16K06768, and No.18H04506.  A part of this work was carried out under the Cooperative Research Project Program of the Research Institute of Electronics, Shizuoka University. SJ is grateful for sabbatical stay supported by the Tokyo Institute of Technology and Shizuoka University. This method of birefringence measurement was developed in conjunction with the proposal M13416 of the Australian synchrotron.}

\subsection*{Additional Information}
The authors declare no competing interests.

%We acknowledge partial support via ARC Discovery DP170100131
%grant. Experiments were carried out via beamtime project No. 12107
%at the Australian Synchrotron IRM Beamline.

\small{J.M. and S.J. come up with the idea of experiments, R.H. developed numerical analysis, wrote the program for image acquisition, and carried out experiments together with M.R.; J.L.L. and V.M. made birefringent samples for testing.  All the authors participated in discussion and analysis of the
results and contributed to editing of the manuscript.}
%\newpage
%\section*{\small REFERENCES}
%\bibliographystyle{pra}   %>>>> bibliography data in report.bib
%\bibliographystyle{version3}
%\bibliographystyle{unsrt,plain,abbrv,alpha}
%\bibliographystyle{aps,apsrmp,apsrev}
%\bibliographystyle{elsart-num}
%\bibliographystyle{achemso}
%\bibliographystyle{spiebib}
\small
%\bibliography{abrio1,paper6b}

%\newpage\newpage
%\vspace{20cm}
\renewcommand\thefigure{\thesection.\arabic{figure}}
\setcounter{figure}{0}

%\newpage
\section{Supplement}

Analytical expressions for the coefficients $A = a_1+ia_2$ and $B
= b_1+ib_2$ of Eqn.~\ref{e4} are following:
%\lipsum[1]

\label{e5}
\begin{eqnarray*}
 a_1=&\sin\frac{\phi_s}{2}\sin\frac{\phi_{LC}}{2}(\cos(2\theta_s)\sin(2\theta_{LC}) - \sin(2\theta_s)\cos(2\theta_{LC})) \\
 a_2=&\sin\frac{\phi_s}{2}\cos\frac{\phi_{LC}}{2}\sin(2\theta_{s}) + \cos\frac{\phi_s}{2}\sin\frac{\phi_{LC}}{2}\sin(2\theta_{LC}) \\
 b_1 =& -\sin\frac{\phi_s}{2}\sin\frac{\phi_{LC}}{2}\sin(2\theta_s)\sin(2\theta_{LC}) + \cos\frac{\phi_s}{2}\cos\frac{\phi_{LC}}{2} - \sin\frac{\phi_s}{2}\sin\frac{\phi_{LC}}{2}\cos(2\theta_s)\cos(2\theta_{LC})  \\
 b_2 =& - \cos\frac{\phi_s}{2}\sin\frac{\phi_{LC}}{2}\cos(2\theta_{LC}) - \sin\frac{\phi_s}{2}\cos\frac{\phi_{LC}}{2}\cos(2\theta_s).\\
\end{eqnarray*}

%\lipsum[1]

\end{document}